\documentclass[11pt,a4paper]{article}
\usepackage[margin=2.3cm]{geometry}
\usepackage[T1]{fontenc}
\usepackage{amsmath,amssymb,amsthm,mathrsfs,bm}
\usepackage{newtxtext,newtxmath}
\usepackage{graphicx,booktabs,authblk,microtype,url}
\usepackage{placeins}
\usepackage{cite}
\usepackage[colorlinks=true,linkcolor=blue,citecolor=blue,urlcolor=blue]{hyperref}
\DeclareMathSizes{8}{8}{6}{5}
\DeclareMathSizes{9}{9}{6}{5}
\DeclareMathSizes{10}{10}{7}{5}
\DeclareMathSizes{10.95}{10.95}{8}{6}
\DeclareMathSizes{11}{10.95}{8}{6}
\newcommand{\dd}{\mathrm d}
\newcommand{\Op}{\operatorname{Op}_{\!W}}
\newcommand{\eps}{\varepsilon}
\newcommand{\KG}{\mathrm{KG}}
\newcommand{\Tr}{\operatorname{Tr}}
\newcommand{\norm}[1]{\left\|#1\right\|}
\numberwithin{equation}{section}
\title{Minimal coupling before or after quantization: a 2PN comparison in Schwarzschild spacetime}
\author[1,2,3]{Hanjun Zou}
\author[1,2,3]{Jingming He}
\author[1,2,3]{Meng Xu}
\author[1,4]{Zhoujian Cao\thanks{Corresponding author: zjcao@bnu.edu.cn.}}
\affil[1]{School of Fundamental Physics and Mathematical Sciences, Hangzhou Institute for Advanced Study, University of Chinese Academy of Sciences, Hangzhou 310024, China}
\affil[2]{School of Physical Sciences, University of Chinese Academy of Sciences, Beijing 100049, China}
\affil[3]{Institute of Theoretical Physics, Chinese Academy of Sciences, Beijing 100190, China}
\affil[4]{School of Physics and Astronomy, Beijing Normal University, Beijing 100875, China}
\date{}
\begin{document}
\maketitle
\begin{abstract}
In curved spacetime, single-particle quantum dynamics can depend on whether minimal coupling is introduced before or after quantization.
Here, we compare two single-particle prescriptions in a static Schwarzschild exterior. 
The canonical prescription quantizes the positive-energy Hamiltonian after minimal coupling, with an operator-ordering ambiguity between the lapse and the spatial energy root.
The Klein--Gordon construction introduces the flat-spacetime squared-energy operator before minimal coupling of its covariant wave equation, then takes the positive spectral root.
Both prescriptions are formulated with the same classical Hamiltonian, scalar product, time normalization and boundary to isolate their differences directly. Their Newtonian and first post-Newtonian terms agree, while the local second post-Newtonian~(2PN) difference is a tidal double commutator. The 2PN level shifts change spectral gaps and rule out unitary equivalence up to a common energy shift for the retained radial spectral series, and unexpanded finite-dimensional root spectra provide a separate numerical check. More generally, the $j$PN differences ($j\geq2$) require kinetic coefficients and, for matched simple levels, state corrections only through order $j-2$. With the same additive optical potential, the vertical motional frequency difference in a reference $^{88}\mathrm{Sr}$ tweezer is about $-1.9\times10^{-54}\,\mathrm{Hz}$ at 2PN order.
\end{abstract}
\section{Introduction}\label{sec:intro}

Understanding how quantum matter evolves in an external gravitational
field is a central problem at the interface of quantum mechanics and
general relativity~\cite{GiuliniGrossardtSchwartz2023}. Modern atomic
interferometers and optical-tweezer platforms are steadily pushing
precision measurements toward regimes sensitive to gravitational
redshifts and spacetime curvature~\cite{Asenbaum2017,CoveyPikovskiBorregaard2025}.
From a theoretical standpoint, however, promoting a relativistic
classical particle to a single-particle quantum theory in curved
spacetime is not unique. In particular, the resulting quantum dynamics
can depend on whether minimal gravitational coupling is introduced
before or after quantization.

In flat spacetime, a free particle of mass $m$ satisfies the mass-shell
relation $E^2=m^2c^4+c^2\bm p^2$. Extending this relation to a static
curved spacetime leads naturally to two established single-particle
routes. The canonical route first minimally couples the classical
mass-shell relation to the spacetime metric and solves for the positive
energy, yielding the classical Hamiltonian in Eq.~\eqref{eq:classical}.
Quantizing the spatial kinetic term on the curved slice via
$K=-\hbar^2\Delta_h$ then introduces a spatial energy root $R$ that does
not commute with the lapse $N$, so an operator ordering must be
specified. DeWitt's foundational analyses of quantum mechanics under
coordinate transformations provide the historical basis for this
canonical formulation~\cite{DeWitt1952,DeWitt1957}. By contrast, the
Klein--Gordon~(KG) route starts from the flat-spacetime squared-energy
operator $m^2c^4-\hbar^2c^2\Delta$, writes the corresponding
second-order wave equation in covariant form, and minimally couples it
to the spacetime metric. On a static background with fixed boundary
conditions, taking the positive spectral root of the resulting
squared-energy operator and matching the scalar product generates the
single-particle dynamics~\cite{Saa1996,GavrilovGitman2000,Mostafazadeh2004}.

Whether these two routes yield equivalent predictions has been examined
primarily through first post-Newtonian~(1PN) order or in flat
accelerated frames. Schwartz and Giulini~\cite{SchwartzGiulini2019}
developed a systematic post-Newtonian~(PN) expansion of the KG equation
and compared it with canonical quantization after matching the scalar
products. Related operator-ordering effects at order $c^{-4}$ have also
been analyzed in Rindler spacetime by Sajnok and
D\k{e}bski~\cite{SajnokDebski2025,SajnokDebskiErratum2026}, who
separated a common energy shift from terms that alter spectral gaps and
heat capacity. However, a Rindler frame describes uniform acceleration
in flat spacetime and does not capture genuine gravitational tidal
gradients. In a curved static spacetime such as the Schwarzschild
exterior, it remains to be determined whether a specified canonical
ordering and the minimally coupled KG reduction remain unitarily
equivalent up to a common energy shift at second post-Newtonian~(2PN)
order, and how curvature-induced ordering differences enter observable
energy gaps.

Here, we address this question by comparing the positive midpoint
canonical prescription~(D) in Eq.~\eqref{eq:D} with the minimally
coupled KG construction in the static Schwarzschild exterior. Both
prescriptions recover the same classical Hamiltonian and are formulated
in a common radial Hilbert space with identical physical parameters,
time normalization and reflecting Dirichlet boundaries. In this common
representation, the difference between the normalized squared energies
takes the exact double-commutator form in Eq.~\eqref{eq:orders_Cdiff}.
The two Hamiltonians share the same Newtonian and 1PN coefficients,
while their leading local difference is the 2PN tidal momentum coupling
in Eq.~\eqref{eq:tidal}. Within the broader symmetric ordering family
$T_\alpha$, varying the lapse ordering adds only a multiplication
potential at 2PN order, so no fixed member reproduces the KG operator
through this order.

Because a unitary transformation and a common energy shift preserve
spectral gaps, we test equivalence using matched radial levels in a
fixed angular sector. The resulting 2PN gap difference in
Eq.~\eqref{eq:gap} rules out unitary equivalence up to a common energy
shift for the retained radial spectral series
(Eq.~\eqref{eq:nounitary}), and unexpanded finite-cavity Galerkin
spectra provide an independent numerical check. More generally, for
$j\geq2$, the $j$PN operator and spectral differences require kinetic
coefficients only through order $j-2$, and for matched simple levels
they also require state corrections only through order $j-2$. As a
physical benchmark, we add the same calibrated optical potential to both
generators and estimate the 2PN vertical motional frequency difference
in a reference $^{88}\mathrm{Sr}$ optical
tweezer~\cite{Norcia2018,Giardini2026}.

The paper is organized as follows. Section~\ref{sec:canonical}
constructs the D and KG generators on the same Schwarzschild domain and
transforms both into the common radial representation.
Section~\ref{sec:kg} compares the two Hamiltonians through 2PN order and
derives the tidal momentum coupling. Section~\ref{sec:spectrum} computes
the 2PN spectral shifts with reflecting boundaries and checks them
against unexpanded finite-cavity spectra. Section~\ref{sec:tweezer}
applies the 2PN tidal difference to a $^{88}\mathrm{Sr}$ optical tweezer
and its Ramsey readout. Section~\ref{sec:discussion} summarizes our
conclusions. Appendix~\ref{app:coeff} lists the explicit radial
differential coefficients at 2PN order, Appendix~\ref{app:isotropic}
verifies the 1PN agreement in isotropic coordinates, and
Appendix~\ref{app:pn_orders} establishes the higher-PN kinetic and state
order bounds.
\section{Common setting and the two quantum prescriptions}\label{sec:canonical}

\subsection{Schwarzschild geometry and reflecting boundaries}

We consider the static Schwarzschild exterior with signature $(-,+,+,+)$~\cite{Wald1984}. The line element is
\begin{equation}
    \dd s^2=-f(r)c^2\dd t^2+f(r)^{-1}\dd r^2+r^2\dd\Omega^2, \qquad f(r)=1-\frac{2\mu}{c^2r},\qquad \mu=GM.\label{eq:geometry}
\end{equation}
Here $M$ is the source mass and $m$ is the test-particle mass. The shift is zero and the lapse is $N=\sqrt f$. The spatial metric $h_{ij}$ is the metric induced on the $t=\mathrm{const}$ slices, and the inverse spatial metric is $h^{ij}$. The positive-energy solution of the minimally coupled classical mass-shell constraint gives the particle Hamiltonian $H_{\rm cl}$~\cite{SchaeferJaranowski2024}, together with the Newtonian potential $\Phi$:
\begin{equation}
    H_{\rm cl}=N\sqrt{m^2c^4+c^2h^{ij}p_ip_j},\qquad \Phi=-\frac{\mu}{r}.\label{eq:classical}
\end{equation}
The momenta $p_i$ are canonically conjugate to the spatial coordinates. Both quantum constructions use the same lapse and spatial metric. The lapse $N$ sets the gravitational redshift of the energy. We require $N(r)\to1$ as $r\to\infty$. With this normalization, $t$ agrees with proper time at infinity. Both constructions use this time coordinate.

We work outside an ideal reflecting sphere. The boundary on each static slice is the two-sphere $r=R_0$. We impose the Dirichlet condition~\cite{Hod2017}
\begin{equation}
    \psi(t,R_0,\theta,\phi)=0.\label{eq:wall}
\end{equation}
For the KG field, the normal current vanishes at the wall. The canonical construction uses the corresponding Dirichlet spatial kinetic operator. Both spectral problems therefore have the same inner boundary condition. We consider either the radial half-line $R_0<r<\infty$ or a finite cavity $R_0<r<R_1$. In the finite cavity, the same Dirichlet condition holds at the outer wall. We discuss the two cases separately. The inner wall must satisfy
\begin{equation}
    R_0>r_{\rm H},\qquad r_{\rm H}=\frac{2\mu}{c^2},\qquad N(R_0)>0.\label{eq:exterior}
\end{equation}
The induced metric on the wall is $-f(R_0)c^2\dd t^2+R_0^2\dd\Omega^2$. The wall is timelike precisely when $R_0>r_{\rm H}$. Static observers there can remain at fixed spatial coordinates. The lapse is positive and the spatial metric is positive definite throughout the domain.

We use $\eps=c^{-2}$ as the PN expansion parameter. We hold $m$, $\mu$, the areal radii and the areal-coordinate canonical momenta fixed. The wall radius $R_0$ and, when present, $R_1$ are fixed at every order. On the scales considered, the weak-field condition is $\mu/(c^2r)\ll1$. The nonrelativistic condition is $h^{ij}p_ip_j/(m^2c^2)\ll1$.

\subsection{The specified canonical prescription}\label{sec:positive}

We describe the particle on a static spatial region $\Omega$ with metric $h_{ij}$ and lapse $N>0$. The volume element is $\dd V_h=\sqrt h\,\dd^3x$, where $h=\det(h_{ij})$. The wavefunction scalar product is
\begin{equation}
    \langle u,v\rangle_h=\int_\Omega\bar u v\,\dd V_h.\label{eq:positive_sample_inner}
\end{equation}
The coordinate momenta can be represented locally as~\cite{Tagirov2003}
\begin{equation}
    p_i=-\mathrm{i}\hbar\left(\partial_i+\frac14\partial_i\log h\right),\qquad [x^i,p_j]=\mathrm{i}\hbar\delta^i_j,\label{eq:positive_sample_momenta}
\end{equation}
where $[X,Y]=XY-YX$ is the commutator. The derivative of $\log h$ compensates for the position dependence of the volume element under integration by parts. We choose the Laplace--Beltrami spatial kinetic operator~\cite{Podolsky1928} with Dirichlet boundary conditions,
\begin{equation}
    Ku=-\hbar^2\Delta_hu =-\frac{\hbar^2}{\sqrt h}\partial_i \left(\sqrt h\,h^{ij}\partial_j u\right), \qquad u|_{\partial\Omega}=0.\label{eq:positive_sample_kinetic}
\end{equation}
We impose the boundary condition of Eq.~\eqref{eq:wall} at each reflecting wall. For smooth wavefunctions that decay sufficiently fast at infinity, integration by parts gives
\begin{align}
    \langle u,Ku\rangle_h={}&\hbar^2\int_\Omega h^{ij}\overline{\partial_i u}\,\partial_j u\,\dd V_h-\hbar^2\int_{\partial\Omega}\bar u\,\partial_nu\,\dd S_h,\label{eq:positive_sample_green}
\end{align}
where $\partial_nu=n^i\partial_i u$ is the derivative along the outward unit normal and $\dd S_h$ is the induced area element. The Dirichlet boundary condition eliminates the surface term and leaves the nonnegative quadratic form
\begin{equation}
    q_K[u]=\hbar^2\int_\Omega h^{ij}\overline{\partial_i u}\,\partial_j u\,\dd V_h \geq0.\label{eq:positive_sample_form}
\end{equation}
The positivity follows from the positive-definite spatial metric. This form defines the nonnegative self-adjoint Dirichlet realization of $K$~\cite{Grigoryan2009}. Its value has units of momentum squared. For a normalized wavefunction, $q_K[u]/(2m)$ is the expectation value of the nonrelativistic kinetic operator $K/(2m)$.

Using this $K$, we define the canonical prescription through the positive spectral root,
\begin{equation}
    R=(m^2c^4+c^2K)^{1/2},\qquad P=N^{1/2},\qquad T_D=PRP,\qquad H_D=T_D-mc^2.\label{eq:D}
\end{equation}
Placing $P$ on both sides of $R$ gives
\begin{equation}
    \langle u,T_Du\rangle_h=\langle Pu,RPu\rangle_h=\norm{R^{1/2}Pu}_h^2\geq0.\label{eq:positive}
\end{equation}
Since $N$ and $R$ do not commute, their product admits different
orderings. Averaging a product with its adjoint gives the family of
symmetric orderings
\[
    T_\alpha=\frac12\left(
    N^\alpha R N^{1-\alpha}+N^{1-\alpha}R N^\alpha
    \right),\qquad \alpha\in\mathbb R .
\]
These orderings have the same classical limit in
Eq.~\eqref{eq:classical}. We choose the midpoint $\alpha=1/2$, at
which the two products coincide and the same factor $P=N^{1/2}$
appears on each side of $R$. The resulting operator $T_D=PRP$ has
the manifestly nonnegative quadratic form in Eq.~\eqref{eq:positive}.
We call this choice the D prescription.

A scalar-curvature coupling replaces $K$ with $K_\xi=-\hbar^2\Delta_h+\xi\hbar^2\,{}^{(3)}R$~\cite{Tagirov2003}. We set $\xi=0$. The static Schwarzschild slice has ${}^{(3)}R=0$~\cite{Anderson2000}, so this curvature term vanishes on the present background.

\subsection{Positive-frequency KG reduction}\label{sec:kg_reduction}

We start from the flat-spacetime squared-energy operator
$m^2c^4-\hbar^2c^2\Delta$, where $\Delta$ is the flat spatial Laplacian.
The corresponding second-order wave equation has the covariant form
$(\Box_\eta-m^2c^2/\hbar^2)\phi=0$. Minimal coupling replaces
$\Box_\eta$ with $\Box_g$ in this equation. We use a static metric
with zero shift,
\begin{equation}
    \dd s^2=-N^2c^2\dd t^2+h_{ij}\dd x^i\dd x^j,
    \qquad \dd V_h=\sqrt h\,\dd^3x,\label{eq:kg_sample_metric}
\end{equation}
where $N>0$ and $h_{ij}$ are independent of $t$. On $\Omega$,
the resulting KG equation
$(\Box_g-m^2c^2/\hbar^2)\phi=0$ is subject to Dirichlet boundary
conditions at the reflecting walls. Isolating the time derivative
gives the squared-energy operator $A$ on the static curved
background~\cite{IshibashiWald2003},
\begin{equation}
    -\hbar^2\partial_t^2\phi=A\phi,\quad
    A=N^2(m^2c^4-\hbar^2c^2\Delta_S),
    \quad
    \Delta_S=\frac1{N\sqrt h}\partial_i(N\sqrt h\,h^{ij}\partial_j).
    \label{eq:Araw}
\end{equation}
We use the conserved KG product~\cite{BarbadoBaezCamargoFuentes2020,BarbadoBaezCamargoFuentes2021}
\begin{equation}
    (\phi_1,\phi_2)_{\rm KG}
    =\frac{\mathrm{i}\hbar}{2mc^2}\int_\Omega\left(\bar\phi_1\partial_t\phi_2-\partial_t\bar\phi_1\,\phi_2\right)\frac{\dd V_h}{N}.\label{eq:kg_sample_product}
\end{equation}
We also define the weighted inner product $\langle u,v\rangle_w=\int_\Omega\bar u v\,\dd V_h/N$.

Restricting to the positive-frequency sector replaces the second-order time equation with
\begin{equation}
    \mathrm{i}\hbar\partial_t\phi=A^{1/2}\phi,\label{eq:kg_sample_frequency}
\end{equation}
where $A^{1/2}$ is the positive spectral root. Substituting this relation for both fields in the KG product gives
\begin{align}
    (\phi_1,\phi_2)_+&=\frac1{2mc^2}\left[\langle\phi_1,A^{1/2}\phi_2\rangle_w+\langle A^{1/2}\phi_1,\phi_2\rangle_w\right]\nonumber\\
    &=\frac1{mc^2}\langle A^{1/4}\phi_1,A^{1/4}\phi_2\rangle_w.\label{eq:kg_sample_positive_product}
\end{align}
The second equality follows from the self-adjointness of $A^{1/2}$. The positive-frequency KG product is therefore positive definite. To rewrite this inner product with the geometric volume element $\dd V_h$, we define
\begin{equation}
    \psi_\KG=\frac1{\sqrt{mc^2}}N^{-1/2}A^{1/4}\phi,\label{eq:normKG}
\end{equation}
which gives
\begin{equation}
    (\phi_1,\phi_2)_+=\frac1{mc^2}\int_\Omega\overline{A^{1/4}\phi_1}\,A^{1/4}\phi_2\,\frac{\dd V_h}{N}
    =\int_\Omega\bar\psi_{\KG,1}\psi_{\KG,2}\,\dd V_h.\label{eq:kg_sample_isometry}
\end{equation}
The factor $A^{1/4}$ absorbs the energy weight of the KG product~\cite{Gerard2019}, and $N^{-1/2}$ absorbs the lapse weight. Normalizing the states removes any dependence on the overall constant in Eq.~\eqref{eq:kg_sample_product}.

On the static background, the transformation in Eq.~\eqref{eq:normKG} is independent of $t$, and $A^{1/4}$ commutes with $A^{1/2}$. The normalized wavefunction therefore evolves as
\begin{equation}
    \mathrm{i}\hbar\partial_t\psi_\KG =N^{-1/2}A^{1/2}N^{1/2}\psi_\KG=B_\KG^{1/2}\psi_\KG,
    \qquad B_\KG=N^{-1/2}AN^{1/2}.\label{eq:kg_sample_evolution}
\end{equation}
The total and rest-subtracted single-particle energy operators are
\begin{equation}
    T_\KG=B_\KG^{1/2},\qquad H_\KG=T_\KG-mc^2.\label{eq:kg_sample_H}
\end{equation}
This defines the normalized positive-frequency single-particle dynamics on the static background~\cite{SchwartzGiulini2019,Mostafazadeh2004,FalconeConti2023}.

To compare with Eq.~\eqref{eq:D}, we express the squared-energy operator in terms of $P$ and $R$. Equations~\eqref{eq:Araw} and~\eqref{eq:kg_sample_evolution} give
\[P^{-1}B_\KG P^{-1}=N(m^2c^4-\hbar^2c^2\Delta_S).\]
In the Schwarzschild vacuum exterior, $\Delta_hN=0$~\cite{Anderson2000}. The product rule gives $N\Delta_Su=[\Delta_h(Nu)+N\Delta_hu]/2$. This yields
\begin{equation}
    B_\KG=\frac12 P\{R^2,N\}P,\label{eq:kg_squared_ordering}
\end{equation}
where $\{X,Y\}=XY+YX$ is the anticommutator.

\subsection{Common radial representation}

We now pass to a radial representation that absorbs the geometric volume factor into the wavefunction. The Schwarzschild spatial metric and lapse are
\begin{equation}
    h_{ij}\dd x^i\dd x^j=f^{-1}\dd r^2+r^2\dd\Omega^2,
    \qquad N=\sqrt f,\qquad f=1-\frac{2\mu}{c^2r},\qquad \mu=GM,
    \label{eq:radial_sample_geometry}
\end{equation}
with volume element $\dd V_h=r^2N^{-1}\dd r\,\dd\Omega$. By spherical symmetry, we separate variables as $\psi=\psi_\ell(r)Y_{\ell q}(\theta,\phi)$, where $\int|Y_{\ell q}|^2\dd\Omega=1$ and $-\Delta_{S^2}Y_{\ell q}=L Y_{\ell q}$ with $L=\ell(\ell+1)$. Here $q$ is the magnetic quantum number and $m$ is the particle mass. For a normalized state in this angular sector,
\begin{equation}
    1=\int_\Omega|\psi|^2\dd V_h=\int_I\frac{r^2}{N}|\psi_\ell(r)|^2\dd r=\int_I|u(r)|^2\dd r,
    \qquad u=rN^{-1/2}\psi_\ell,\label{eq:radial_sample_norm}
\end{equation}
where $I=(R_0,\infty)$ or $I=(R_0,R_1)$ is the radial domain. Here $|u|^2\dd r$ is the radial probability measure, with the geometric weight absorbed into $u$.

In a fixed angular sector, the three-dimensional operator $K=-\hbar^2\Delta_h$ defines the radial operator $K_\ell$ through
\begin{equation}
\begin{gathered}
    K\left[\psi_\ell(r)Y_{\ell q}(\theta,\phi)\right]=\left[K_\ell\psi_\ell(r)\right]Y_{\ell q}(\theta,\phi),\\[5pt]
    K_\ell\psi_\ell=-\hbar^2\frac{N}{r^2}\partial_r\left(r^2N\partial_r\psi_\ell\right)+\frac{\hbar^2L}{r^2}\psi_\ell.
\end{gathered}\label{eq:radial_sample_Kell}
\end{equation}
The kinetic operator acting on the normalized radial amplitude $u$ is
\begin{equation}
    K_{\mathrm{rad}} u=\frac{r}{N^{1/2}}\,K_\ell\!\left(\frac{N^{1/2}}{r}\,u\right).\label{eq:radial_operator_map}
\end{equation}
The subscript $\mathrm{rad}$ denotes operators in $L^2(I,\dd r)$. Using $N=\sqrt f$, we obtain
\begin{align}
    K_{\mathrm{rad}} u
    &=-\hbar^2\frac{N^{1/2}}{r}\,\partial_r\!\left[r^2N\,\partial_r\!\left(\frac{N^{1/2}}{r}u\right)\right]+\frac{\hbar^2L}{r^2}u\nonumber\\
    &=-\hbar^2\partial_r(f\partial_r u)+\hbar^2\left[\frac{L}{r^2}+\frac{f'}{2r}-\frac{f''}{4}+\frac{(f')^2}{16f}\right]u, \label{eq:radial_sample_transport}
\end{align}
where primes denote derivatives with respect to $r$. Substituting $f=1-2\mu\eps/r$ with $\eps=c^{-2}$ gives the exact radial operator
\begin{equation}
    K_{\mathrm{rad}}=-\hbar^2\partial_r(f\partial_r)+\hbar^2\left[\frac{L}{r^2}+\frac{2\mu\eps}{r^3}+\frac{\mu^2\eps^2}{4r^4f}\right].\label{eq:Krad}
\end{equation}
The potential-like radial terms arise from the spatial metric and the radial rescaling in Eq.~\eqref{eq:radial_sample_norm}.

The radial scalar product is $\int_I\bar u v\,\dd r$. Since $R_0>2\mu/c^2$, the factor $rN^{-1/2}$ is finite and nonzero at each wall. The boundary condition $\psi_\ell(R_i)=0$ is therefore equivalent to $u(R_i)=0$ at each boundary. This unitary transformation preserves both the Dirichlet boundary conditions and the operator spectra.

In a fixed angular sector, let $B_{\KG,\ell}$ be the radial part of $B_\KG$. On the normalized radial amplitude $\psi_\KG$, the squared-energy operator is
\begin{equation}
    B_{\mathrm{rad}}\psi_\KG=\frac{r}{N^{1/2}}\,B_{\KG,\ell}\!\left(\frac{N^{1/2}}{r}\,\psi_\KG\right).\label{eq:kg_sample_radial_B}
\end{equation}
With $N^2=f=1-2\mu\eps/r$, $\eps=c^{-2}$ and $L=\ell(\ell+1)$, the radial part of the KG spatial operator $A$ is~\cite{Hod2017}
\begin{equation}
    A_\ell=f\left[m^2c^4-\frac{\hbar^2c^2}{r^2}\partial_r(r^2f\partial_r)+\frac{\hbar^2c^2L}{r^2}\right].\label{eq:kg_sample_Aradial}
\end{equation}
Substituting $B_{\KG,\ell}=N^{-1/2}A_\ell N^{1/2}$ into Eq.~\eqref{eq:kg_sample_radial_B} and differentiating gives
\begin{align}
    B_{\mathrm{rad}}\psi_\KG
    &=m^2c^4f\psi_\KG -\hbar^2c^2\frac{N}{r}\partial_r\!\left[r^2f\partial_r\!\left(\frac Nr\psi_\KG\right)\right]+\frac{\hbar^2c^2fL}{r^2}\psi_\KG\nonumber\\
    &=m^2c^4f\psi_\KG -\hbar^2c^2\partial_r(f^2\partial_r\psi_\KG)
    +\hbar^2c^2\left[\frac{fL}{r^2}+\frac{ff'}r-\frac{ff''}{2}-\frac{(f')^2}{4}\right]\psi_\KG,\label{eq:kg_sample_Bderivation}
\end{align}
where primes denote derivatives with respect to $r$. Substituting $f'=2\mu\eps/r^2$ and $f''=-4\mu\eps/r^3$ gives the exact radial operator
\begin{equation}
\begin{aligned}
    B_{\mathrm{rad}}=m^2c^4f-\hbar^2c^2\partial_r(f^2\partial_r)+\hbar^2c^2\left[\frac{fL}{r^2}+\frac{4\mu\eps f}{r^3}-\frac{\mu^2\eps^2}{r^4}\right].
\end{aligned}\label{eq:Cexact}
\end{equation}

Equations~\eqref{eq:D} and~\eqref{eq:kg_sample_H} give the two rest-subtracted Hamiltonians in the common radial space \(L^2(I,\dd r)\). For the D prescription,
\begin{equation}
    H_{D,\mathrm{rad}}=PR_{\mathrm{rad}}P-mc^2,\label{eq:agreement_sample_H}
\end{equation}
where
\[P=N^{1/2},\qquad R_{\mathrm{rad}}=(m^2c^4+c^2K_{\mathrm{rad}})^{1/2}.\]
The radial KG Hamiltonian is
\begin{equation}
    H_{\KG,\mathrm{rad}}=B_{\mathrm{rad}}^{1/2}-mc^2,\label{eq:agreement_sample_HKG}
\end{equation}
where $B_{\mathrm{rad}}$ is given by Eq.~\eqref{eq:Cexact}.
\section{Post-Newtonian Hamiltonians through 2PN}\label{sec:kg}

\subsection{Normalized squared energies}\label{sec:squared_energy}

For the two radial Hamiltonians in Eqs.~\eqref{eq:agreement_sample_H} and~\eqref{eq:agreement_sample_HKG}, write $T_X=mc^2+H_X$ with $X=D,\KG$. We suppress the radial subscript where no ambiguity arises. The normalized squared energy and the inverse positive root are
\begin{equation}
\begin{aligned}
    C_X&=\frac{T_X^2-m^2c^4}{2mc^2}=H_X+\frac{\eps}{2m}H_X^2,\\
    H_X&=mc^2\left(\sqrt{1+\frac{2C_X}{mc^2}}-1\right).
\end{aligned}\label{eq:agreement_sample_Cdefinition}
\end{equation}
In the static vacuum exterior, Eqs.~\eqref{eq:agreement_sample_H} and~\eqref{eq:kg_squared_ordering} give $T_D^2=PRNRP$ and $T_\KG^2=B_{\mathrm{rad}}=P\{R^2,N\}P/2$. For KG, the normalization gives
\begin{equation}
    C_\KG=\frac{m}{2\eps}(N^2-1)+\frac{1}{4m}P\{K,N\}P .\label{eq:orders_CKG}
\end{equation}
In the fixed areal-radius expansion, $N^2=1+2\eps\Phi$ exactly.

Write $S=R-mc^2$. The identity $RNR-\{R^2,N\}/2=-[R,[R,N]]/2$ then yields
\begin{equation}
    C_D-C_\KG=-\frac{\eps}{4m}P[S,[S,N]]P .\label{eq:orders_Cdiff}
\end{equation}
In the formal local PN expansions below, we evaluate the differential-operator coefficients on $C_c^\infty(I)$, the space of smooth radial test functions with compact support in the interior of $I$.

\subsection{Common Newtonian and 1PN terms}

At fixed areal radius, Eq.~\eqref{eq:Cexact} gives the exact polynomial
\begin{equation}
    C_\KG=\frac{B_{\mathrm{rad}}-m^2c^4}{2mc^2}
    =H_0+\eps C_{\KG,1}+\eps^2 C_{\KG,2},\label{eq:kg_sample_Bexpansion}
\end{equation}
where, with $\Phi=-\mu/r$,
\begin{equation}
\begin{aligned}
    H_0&=\frac{K_0}{2m}+m\Phi,\\[3pt]
    C_{\KG,1}&=\frac{2\hbar^2\mu}{m}\partial_r(r^{-1}\partial_r)+\frac{\hbar^2\mu(2-L)}{mr^3},\\[3pt]
    C_{\KG,2}&=-\frac{2\hbar^2\mu^2}{m}\partial_r(r^{-2}\partial_r)-\frac{9\hbar^2\mu^2}{2mr^4}.
\end{aligned}\label{eq:Ccoeff}
\end{equation}

To expand $C_D$ using Eq.~\eqref{eq:orders_Cdiff}, write $K_{\mathrm{rad}}=K_0+\eps K_1+\eps^2K_2+\cdots$, with
\begin{equation}
\begin{aligned}
    K_0&=-\hbar^2\partial_r^2+\frac{\hbar^2L}{r^2},\\[3pt]
    K_1&=2\hbar^2\mu\partial_r(r^{-1}\partial_r)+\frac{2\hbar^2\mu}{r^3},\\[3pt]
    K_2&=\frac{\hbar^2\mu^2}{4r^4}.
\end{aligned}\label{eq:Kcoeff}
\end{equation}
The lapse and spatial root expand as
\begin{equation}
\begin{aligned}
    N&=1+\eps\Phi-\frac{\eps^2}{2}\Phi^2+\cdots,\\[3pt]
    P&=1+\frac{\eps}{2}\Phi+\cdots,\\[3pt]
    R_{\mathrm{rad}}&=mc^2+\frac{K_{\mathrm{rad}}}{2m}-\frac{\eps K_{\mathrm{rad}}^2}{8m^3}+\cdots.
\end{aligned}\label{eq:agreement_sample_Dexpansion}
\end{equation}
Inserting these into Eq.~\eqref{eq:orders_Cdiff}, the first contribution comes from $[S,N]=\eps[K_0,\Phi]/(2m)+\cdots$. Together with Eq.~\eqref{eq:kg_sample_Bexpansion}, this gives
\begin{equation}
\begin{aligned}
    C_D&=H_0+\eps C_{D,1}+\eps^2 C_{D,2}+\cdots,\\
    C_{D,1}&=C_{\KG,1},\qquad
    C_{D,2}=C_{\KG,2}-\frac{1}{16m^3}[K_0,[K_0,\Phi]] .
\end{aligned}\label{eq:CDexpansion}
\end{equation}

Expanding the common positive spectral root, as in the Foldy--Wouthuysen representation for scalar particles~\cite{AcciolyBlas2002,Silenko2013}, gives
\begin{equation}
    H_X=C_X-\frac{\eps}{2m}C_X^2+\frac{\eps^2}{2m^2}C_X^3+\cdots .\label{eq:Croot}
\end{equation}
The first-order coefficients are
\begin{equation}
\begin{gathered}
    H_{1,D}=C_{D,1}-\frac{H_0^2}{2m},\qquad
    H_{1,\KG}=C_{\KG,1}-\frac{H_0^2}{2m},\\
    H_{1,D}=H_{1,\KG}\equiv H_1
    =-\frac m2\Phi^2+\frac1{2m}\left(K_1+\frac12\{\Phi,K_0\}\right)-\frac{K_0^2}{8m^3}.
\end{gathered}\label{eq:H1}
\end{equation}
Appendix~\ref{app:isotropic} compares this coefficient with the normalized vacuum parametrized post-Newtonian~(PPN) Hamiltonian of Ref.~\cite{SchwartzGiulini2019}, and Ref.~\cite{HartongEtAl2024} develops a related covariant $1/c^2$ coupling prescription.

\subsection{Complete 2PN coefficients and the tidal difference}\label{sec:tidal_difference}
Equation~\eqref{eq:Croot} gives $H_{2,X}=C_{X,2}-\{H_0,C_{X,1}\}/(2m)+H_0^3/(2m^2)$. The terms involving $H_0$ and $C_{X,1}$ are common. Equation~\eqref{eq:CDexpansion} then gives the local difference
\begin{equation}
    \delta H_2:=H_{2,D}-H_{2,\KG}=C_{D,2}-C_{\KG,2}=-\frac1{16m^3}[K_0,[K_0,\Phi]]. \label{eq:delta}
\end{equation}
Using Eqs.~\eqref{eq:orders_CKG} and~\eqref{eq:Ccoeff} gives the individual coefficients
\begin{align}
    H_{2,D}={}&\frac m2\Phi^3+\frac1{2m}\left(K_2+\frac12\{\Phi,K_1\}-\frac38\{\Phi^2,K_0\}+\frac14\Phi K_0\Phi\right)\nonumber\\
    &-\frac1{8m^3}\left(\{K_0,K_1\}+\frac12\{\Phi,K_0^2\}\right)+\frac{K_0^3}{16m^5},\label{eq:H2D}\\
    H_{2,\KG}={}&C_{\KG,2}-\frac1{2m}\{H_0,C_{\KG,1}\}+\frac{H_0^3}{2m^2}.\label{eq:H2KG}
\end{align}
Appendix~\ref{app:coeff} gives the explicit radial differential coefficients multiplying $u,u',\ldots,u^{(6)}$. The local radial expression for the difference is second order,
\begin{equation}
    \delta H_2u=\frac{\hbar^4\mu}{2m^3r^3}u''-\frac{3\hbar^4\mu}{2m^3r^4}u'+\frac{\hbar^4\mu(L+6)}{4m^3r^5}u . \label{eq:radialdelta}
\end{equation}
In Newtonian Cartesian coordinates, Eq.~\eqref{eq:delta} uses $K_0=-\hbar^2\Delta$. The product rule gives
\[
\begin{aligned}[c]
 [\Delta,[\Delta,\Phi]]\psi
 &= [\Delta,\Delta\Phi]\psi+2[\Delta,\Phi_{,i}\partial_i]\psi\\
 &= (\Delta^2\Phi)\psi+4(\partial_i\Delta\Phi)\partial_i\psi
    +4\Phi_{,ij}\partial_i\partial_j\psi .
\end{aligned}
\]
In the vacuum exterior, $\Delta\Phi=0$, so the last term gives
\begin{equation}
    \frac{\delta H_2}{c^4}\psi
    =\frac{\hbar^2}{4m^3c^4}\,\Phi_{,ij}(x)\hat p_i\hat p_j\psi,\label{eq:tidal}
\end{equation}
where $\hat p_i=-\mathrm{i}\hbar\partial_i$. The Hessian measures tidal gravity~\cite{Poisson2009,PoissonWill2014}. The resulting coupling vanishes for a uniform linear Newtonian potential and is nonzero in the Schwarzschild exterior. The term is quantum and vanishes with $\hbar$. It arises because the lapse does not commute with the spatial kinetic root.

With $b=\alpha-1/2$, Appendix~\ref{app:pn_kinetic} gives
\[
    T_\alpha-T_D
    =-\frac{b^2\hbar^2}{2mc^4}|\nabla\Phi|^2+O(c^{-6}).
\]
For every fixed $\alpha$, $T_\alpha$ and $T_{\KG}$ agree through 1PN and differ at 2PN.

At fixed lapse expansion, the general construction of an individual $j$PN Hamiltonian coefficient can involve $K_j$.
The explicit factor of $\eps$ in Eq.~\eqref{eq:orders_Cdiff} and the fact that $[S,N]=O(\eps)$
limit the kinetic input for the $j$PN differences of both $C_X$ and $H_X$
to coefficients through $K_{j-2}$ for $j\geq2$, as shown in Appendix~\ref{app:pn_kinetic}.
\section{The 2PN spectrum}\label{sec:spectrum}

A time-independent unitary transformation $V$ acts simultaneously as $H\mapsto VHV^\dagger$, $\rho\mapsto V\rho V^\dagger$ and $F\mapsto VFV^\dagger$. When the operator domains and boundary conditions transform together, $V$ preserves spectral gaps and Born probabilities $\Tr(\rho F)$~\cite{ReedSimon1980,SchwartzGiulini2019}. Physical predictions depend on the unitary equivalence class of the Hamiltonian rather than on its explicit operator form, just as observables in effective field theories are invariant under field redefinitions~\cite{CriadoPerezVictoria2019}. We test whether the D and KG prescriptions are unitarily equivalent through 2PN order. This comparison fixes the time normalization, the physical parameters and the reflecting boundaries.

\subsection{2PN level shifts and spectral gaps}

Fix an angular sector and a magnetic component. Let $u_n$ be a normalized eigenstate of the common Newtonian radial Hamiltonian $H_0=K_0/(2m)+m\Phi$ with a simple, isolated eigenvalue $e_n^{(0)}$. Both prescriptions obey the Dirichlet conditions of Section~\ref{sec:canonical}. Define the quadratic forms $c_X$ by
\[
c_X[u,v]=\frac{\eps}{2m}\bigl(\langle T_Xu,T_Xv\rangle-m^2c^4\langle u,v\rangle\bigr).
\]
Equations~\eqref{eq:kg_sample_Bexpansion} and~\eqref{eq:CDexpansion} give the common coefficients $H_0$ and $C_{D,1}=C_{\KG,1}$ of $c_D$ and $c_\KG$. For matched branches admitting a 2PN expansion, this gives the same leading and first-order energy coefficients. Write
\begin{equation}
    E_{X,n}=e_n^{(0)}+\eps e_n^{(1)}+\eps^2 e_{X,n}^{(2)}+\cdots.\label{eq:energyseries}
\end{equation}
\begingroup
\postdisplaypenalty=10000
The energy eigenstate $u_{X,n}$ satisfies
\begin{equation}
    T_{X,\mathrm{rad}}^2u_{X,n}=(mc^2+E_{X,n})^2u_{X,n},
\end{equation}
where $u_{X,n}=u_n+\eps v_n+\eps^2\chi_{X,n}+\cdots$.\par
\endgroup
The eigenvalue of $C_{X,\mathrm{rad}}$ on this state is therefore
\begin{equation}
\begin{aligned}
    \lambda_{X,n}
    &=\frac{(mc^2+E_{X,n})^2-m^2c^4}{2mc^2}=E_{X,n}+\frac{\eps}{2m}E_{X,n}^2\\
    &=e_n^{(0)}+\eps\left[e_n^{(1)}+\frac{(e_n^{(0)})^2}{2m}\right]
    +\eps^2\left[e_{X,n}^{(2)}+\frac{e_n^{(0)}e_n^{(1)}}{m}\right]+O(\eps^3).
\end{aligned}\label{eq:squared_energy_eigenvalue}
\end{equation}
The common terms cancel, giving
\begin{equation}
    \lambda_{D,n}-\lambda_{\KG,n}=\eps^2\bigl(e_{D,n}^{(2)}-e_{\KG,n}^{(2)}\bigr)+O(\eps^3).
\end{equation}

For a test state $\eta$, the eigenvalue equation reads $c_X[\eta,u_{X,n}]=\lambda_{X,n}\langle\eta,u_{X,n}\rangle$. Write $\lambda_{X,n}^{(j)}=[\eps^j]\lambda_{X,n}$ and $c_{X,j}=[\eps^j]c_X$. With $\langle u_n,v_n\rangle=0$, the first-order equation is
\begin{equation}
    (e_n^{(0)}-H_0)v_n=(C_{\KG,1}-\lambda_n^{(1)})u_n,\qquad \lambda_n^{(1)}=\langle u_n,C_{\KG,1}u_n\rangle .
\end{equation}
The simple isolated Newtonian level fixes the same $v_n$ on the orthogonal complement of $u_n$ for both prescriptions~\cite{Kato1995}.
Testing the second-order equation with $u_n$ gives
\begin{align}
    \lambda_{X,n}^{(2)}&=c_{X,2}[u_n,u_n]+c_1[u_n,v_n]
    +\langle (H_0-e_n^{(0)})u_n,\chi_{X,n}\rangle
    \nonumber\\
    &=c_{X,2}[u_n,u_n]+c_1[u_n,v_n],
\end{align}
where $c_1=c_{D,1}=c_{\KG,1}$. The common iteration term cancels. The equality of the 2PN differences in $\lambda_{X,n}$ and $E_{X,n}$ then gives
\begin{equation}
    e_{D,n}^{(2)}-e_{\KG,n}^{(2)}
    =\lambda_{D,n}^{(2)}-\lambda_{\KG,n}^{(2)}
    =[\eps^2](c_D-c_\KG)[u_n,u_n].\label{eq:diagonal_level_difference}
\end{equation}

In the radial representation, Eq.~\eqref{eq:kg_squared_ordering} gives $B_{\mathrm{rad}}=P\{R_{\mathrm{rad}}^2,N\}P/2$. Let $z=Pu$. We use the expansions in Eqs.~\eqref{eq:Kcoeff} and~\eqref{eq:agreement_sample_Dexpansion} and $\operatorname{Re}\langle z,[R_{\mathrm{rad}},N]z\rangle=0$ to obtain
\begin{equation}
\begin{aligned}
    \|T_{D,\mathrm{rad}}u\|^2-\|T_{\KG,\mathrm{rad}}u\|^2
    &=\langle R_{\mathrm{rad}}z,NR_{\mathrm{rad}}z\rangle-\operatorname{Re}\langle R_{\mathrm{rad}}z,R_{\mathrm{rad}}(Nz)\rangle\\
    &=-\operatorname{Re}\langle R_{\mathrm{rad}}z,[R_{\mathrm{rad}},N]z\rangle\\
    &=-\frac{\eps}{4m^2}\operatorname{Re}\langle K_0u,[K_0,\Phi]u\rangle+\cdots .
\end{aligned}\label{eq:squared_energy_norm_difference}
\end{equation}
Hence, the coefficient of $\eps^2$ in $c_D-c_\KG$ is
\begin{equation}
    [\eps^2](c_D-c_\KG)[u,u]=-\frac{\mathfrak C_\Phi[u]}{16m^3},\label{eq:wall_alt_coeff}
\end{equation}
where
\begin{equation}
    \mathfrak C_\Phi[u]=2\operatorname{Re}\langle K_0u,w\rangle,\qquad w=[K_0,\Phi]u.\label{eq:weakcomm}
\end{equation}
Using $K_0=-\hbar^2\partial_r^2+\hbar^2L/r^2$ from Eq.~\eqref{eq:Kcoeff}, the commutator $\mathcal A_\Phi=[K_0,\Phi]$ acts as
\begin{align}
    \mathcal A_\Phi&=-\hbar^2(2\Phi'\partial_r+\Phi''), \label{eq:wall_alt_A}\\
    \mathcal A_\Phi u&=w=-\hbar^2(\Phi''u+2\Phi'u').
\end{align}
The Newtonian eigenvalue equation gives
\begin{equation}
    K_0u_n=2m(e_n^{(0)}-m\Phi)u_n=q_nu_n.\label{eq:wall_alt_Ku}
\end{equation}
We substitute this relation into Eq.~\eqref{eq:weakcomm}, apply $2\operatorname{Re}(\bar u_nu_n')=(|u_n|^2)'$, and integrate by parts:
\begin{equation}
\begin{aligned}
    \mathfrak C_\Phi[u_n]
    &=-2\hbar^2\int_I q_n(\Phi'|u_n|^2)'\dd r=2\hbar^2\int_I q_n'\Phi'|u_n|^2\dd r =-4m^2\hbar^2\int_I|\Phi'|^2|u_n|^2\dd r .
\end{aligned}
\end{equation}
With $\Phi=-\mu/r$, this gives
\begin{equation}
    \mathfrak C_\Phi[u_n]=-4m^2\hbar^2G_n,\qquad G_n=\langle u_n,|\Phi'|^2u_n\rangle=\mu^2\langle r^{-4}\rangle_n .\label{eq:Gn}
\end{equation}
The 2PN level difference is therefore
\begin{equation}
    [E_{D,n}-E_{\KG,n}]_{2\mathrm{PN}}=\frac{\hbar^2G_n}{4mc^4}.\label{eq:leveldifference}
\end{equation}

For two fixed levels, Eq.~\eqref{eq:leveldifference} gives the 2PN gap difference
\begin{equation}
    [(E_{D,n}-E_{D,k})-(E_{\KG,n}-E_{\KG,k})]_{2\mathrm{PN}}=\frac{\hbar^2}{4mc^4}(G_n-G_k). \label{eq:gap}
\end{equation}
On the half-line, every bound state satisfies $G_n>0$, and $G_n\to0$ as $n\to\infty$. Consequently, $G_n\ne G_k$ for some pair of levels~\cite{BetheSalpeter1957,Baumann2019}. Unitary transformations and common energy shifts preserve spectral gaps. Equation~\eqref{eq:gap} therefore rules out
\begin{equation}
    H_{D,\mathrm{rad}}\stackrel{2\mathrm{PN}}{=}U H_{\KG,\mathrm{rad}} U^\dagger+b(c)I\label{eq:nounitary}
\end{equation}
for the matched perturbative branches in the fixed radial sector.

For matched simple Dirichlet levels admitting PN expansions, the $j$PN energy difference
for $j\geq2$ likewise requires kinetic coefficients only through $K_{j-2}$
and state corrections only through order $j-2$ (Appendix~\ref{app:pn_spectral}).

\subsection{Numerical check in a finite cavity}\label{sec:numerical}

We set $m=\hbar=\mu=1$ and $\ell=2$ in the cavity $I=[R_0,R_1]=[1,40]$ with Dirichlet conditions at both walls. We construct an orthonormal Galerkin basis $\{\varphi_i\}$ from $P_j(z)-P_{j+2}(z)$, where $P_j$ are Legendre polynomials and $z=2(r-R_0)/(R_1-R_0)-1$~\cite{Shen1994,ShenTangWang2011}. The spectra reported below use 56 basis functions, 336 Gauss--Legendre quadrature nodes and 55-digit arithmetic. The multiplication and derivative matrices are
\begin{equation}
\begin{aligned}
    M(v)_{ij}&=\int_I \bar\varphi_i v\varphi_j\,\dd r,\\
    S(a)_{ij}&=-\int_I \bar\varphi_i(a\varphi_j')'\,\dd r\\
    &=-[\bar\varphi_i a\varphi_j']_{R_0}^{R_1}+\int_I a\,\overline{\varphi_i'}\varphi_j'\,\dd r=\int_I a\,\overline{\varphi_i'}\varphi_j'\,\dd r .
\end{aligned}
\end{equation}
Set $b=\mu/c^2$, $f=1-2b/r$ and $N=\sqrt f$. Equations~\eqref{eq:Krad} and~\eqref{eq:Cexact} then give the radial kinetic matrix $\widehat K_{\mathrm{rad}}$ and the shifted KG squared-energy matrix $\widehat B_P$:
\begin{equation}
\begin{aligned}
    \widehat K_{\mathrm{rad}}&=S(f)+M\left(\frac{L}{r^2}+\frac{2b}{r^3}+\frac{b^2}{4r^4f}\right),\\
    \widehat B_P&=c^2\left[S(f^2)+M\left(\frac{fL}{r^2}+\frac{4bf}{r^3}-\frac{b^2}{r^4}-\frac{2\mu}{r}\right)\right].
\end{aligned}
\end{equation}
With $m=1$, the matrix Hamiltonian corresponding to Eq.~\eqref{eq:agreement_sample_H} is
\begin{equation}
    \widehat H_{D,\mathrm{rad}}=M(N)^{1/2}\sqrt{c^4\mathbb I+c^2\widehat K_{\mathrm{rad}}}\,M(N)^{1/2}-c^2\mathbb I.
\end{equation}
For KG, Eq.~\eqref{eq:agreement_sample_HKG} gives
\begin{equation}
    \widehat H_{\KG,\mathrm{rad}}=\sqrt{c^4\mathbb I+\widehat B_P}-c^2\mathbb I,
\end{equation}
where $\mathbb I$ is the identity matrix. For the three lowest radial levels, we compare the scaled difference $c^4(E_{D,n}-E_{\KG,n})$ with the analytical 2PN coefficient $G_n/4$ from Eq.~\eqref{eq:leveldifference}.

\begin{figure}[htbp]
    \centering
    \includegraphics[width=0.67\linewidth]{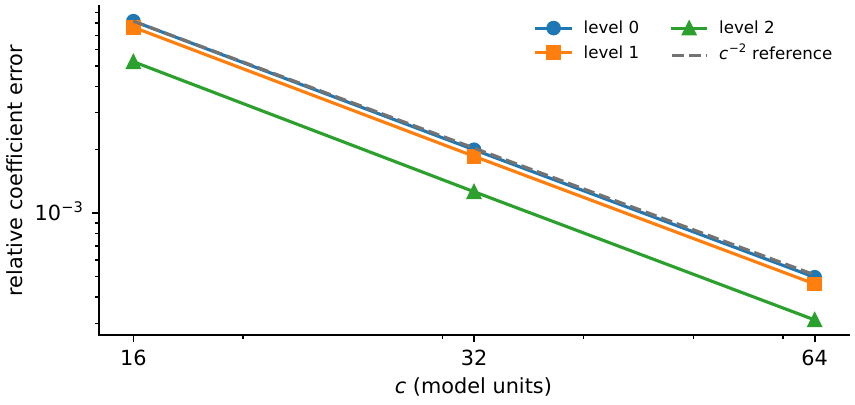}
    \caption{Relative error $\lvert (E_{D,n}-E_{\KG,n})-G_n/(4c^4)\rvert/\lvert G_n/(4c^4)\rvert$ with respect to the analytical 2PN level difference. The dashed line shows the $c^{-2}$ reference.}\label{fig:convergence}
\end{figure}

Table~\ref{tab:spectrum} compares the results at $c=64$. The largest relative deviation is about $5.0\times10^{-4}$. Across $c=16,32,64$, the deviations in Fig.~\ref{fig:convergence} scale approximately as $c^{-2}$, which matches the next PN order. At $c=32$, increasing the basis from 56 to 72 functions and the quadrature from 336 to 432 nodes changes each scaled level difference by less than $10^{-5}$ in relative terms.
\begin{table}[htbp]
    \centering
    \caption{Analytical and numerical coefficients for the three lowest radial levels at $c=64$, in units of $10^{-5}$.}\label{tab:spectrum}
    \begin{tabular}{ccc}
    \toprule
    Level $n$ & Analytical 2PN coefficient $G_n/4$ & Unexpanded $c^4(E_{D,n}-E_{\KG,n})$\\
    \midrule
    0 &12.2834254&12.2895298\\
    1 &6.1602981&6.1631422\\
    2 &7.2703187&7.2725855\\
    \bottomrule
    \end{tabular}
\end{table}
\FloatBarrier

\section{2PN motional frequency difference in a strontium optical tweezer}\label{sec:tweezer}
The radial spectra in Section~\ref{sec:spectrum} show how the 2PN ordering difference changes energy gaps at fixed reflecting boundaries. Precision atomic experiments provide a natural arena to probe how spacetime curvature and gravity gradients affect quantum phases and transition frequencies~\cite{Asenbaum2017,CoveyPikovskiBorregaard2025}. In an optical tweezer, a smooth potential confines the atom, and the local tidal operator~\eqref{eq:tidal} shifts the frequency of a motional transition. For a single $^{88}\mathrm{Sr}$ atom in the same optical potential under both prescriptions, we estimate the motional frequency difference and the corresponding phase difference in an ideal Ramsey measurement.

\subsection{Additive optical coupling and the motional shift}\label{sec:coupling}
A tightly focused beam forms an optical dipole trap~\cite{Grimm2000}. In the strontium platform of Giardini et al., an objective with high numerical aperture focuses the trapping beam onto the atoms in the vacuum chamber. The same objective collects fluorescence to detect single atoms~\cite{Giardini2026}. We consider one $^{88}\mathrm{Sr}$ atom of mass $m$ in the electronic ground state $^{1}S_0$~\cite{Norcia2018}. The spatially varying AC Stark shift from the off-resonant trapping beam gives the optical dipole potential $U(\bm x)$ for the center-of-mass motion. Through 2PN order we use the same independently calibrated potential and static time normalization for the two prescriptions,
\begin{equation}
 H_X^{\mathrm{trap}}=H_X^{\mathrm{free}}+U(\bm x).
 \label{eq:trap_coupling}
\end{equation}
Here $H_X^{\mathrm{free}}$ is the rest-subtracted generator in Eq.~\eqref{eq:D} for D or Eq.~\eqref{eq:kg_sample_H} for KG, expressed in the local spatial coordinates.

We place the beam focus at the origin where the beam is narrowest, and denote the atom's position by $\bm x=(x,y,z)$. The $z$ axis points vertically upward and the $x$ and $y$ axes are horizontal, with $y$ along the beam. We model Earth as static and spherical. Let $R_\oplus$ be the distance from Earth's center to the focus and $r$ the distance from Earth's center to the atom. In the local Newtonian Cartesian chart,
\[
 r=\sqrt{(R_\oplus+z)^2+x^2+y^2}.
\]
Expanding this distance about the focus to quadratic order gives
\[
 r-R_\oplus=z+\frac{x^2+y^2}{2R_\oplus}
   +O\!\left(\frac{|\bm x|^3}{R_\oplus^2}\right).
\]
The Newtonian potential per unit mass is $\Phi=-\mu/r$, where $\mu=GM_\oplus$, $G$ is Newton's gravitational constant and $M_\oplus$ is Earth's mass. Taylor expansion in $r-R_\oplus$ gives
\[
 \begin{aligned}
 \Phi
 &=-\frac{\mu}{R_\oplus}
   +\frac{\mu}{R_\oplus^2}(r-R_\oplus)
   -\frac{\mu}{R_\oplus^3}(r-R_\oplus)^2
   +O\!\left(\frac{\mu|r-R_\oplus|^3}{R_\oplus^4}\right)\\
 &=-\frac{\mu}{R_\oplus}
   +\frac{\mu z}{R_\oplus^2}
   +\frac{\mu}{2R_\oplus^3}(x^2+y^2-2z^2)
   +O\!\left(\frac{\mu|\bm x|^3}{R_\oplus^4}\right).
 \end{aligned}
\]
The constant is $\Phi_{\rm ref}=-\mu/R_\oplus$. The coefficient $g=\mu/R_\oplus^2$ is the magnitude of the downward gravitational acceleration at the focus. The quadratic term gives the tidal Hessian,
\begin{equation}
 Q_{ij}=\left.\partial_i\partial_j\Phi\right|_{\bm x=0},
 \qquad Q=\frac{\mu}{R_\oplus^3}\operatorname{diag}(1,1,-2).
 \label{eq:localPhi}
\end{equation}
The indices $i,j\in\{x,y,z\}$ label the local Cartesian directions, and $\partial_i$ denotes differentiation with respect to $x_i$. We write $Q_i=Q_{ii}$ and $q=\mu/R_\oplus^3$, so the horizontal entries are $q$ and the vertical entry is $-2q$.

For an attractive optical dipole trap, the minimum of $U$ is at the intensity maximum of the Gaussian beam. The potential is
\begin{equation}
 U(\bm x)=-U_0\frac{\exp[-2(x^2+z^2)/w^2(y)]}{1+y^2/z_R^2}.
 \label{eq:gaussian_trap}
\end{equation}
Here $U_0>0$ is the optical trap depth and $\lambda$ is the trapping wavelength. The radius $w(y)$ is measured where the intensity falls to $1/e^2$ of its on-axis value. Its value at the focus is the beam waist $w_0$, and $z_R$ is the Rayleigh length,
\[
 w^2(y)=w_0^2(1+y^2/z_R^2),
 \qquad z_R=\frac{\pi w_0^2}{\lambda}.
\]
With $\eta^2=(x^2+z^2)/w_0^2+y^2/z_R^2$, the expansion is
\[
 U=-U_0+\frac{2U_0}{w_0^2}(x^2+z^2)
        +\frac{U_0}{z_R^2}y^2+O(U_0\eta^4).
\]
Writing the quadratic part as $\tfrac12m\sum_i\omega_i^2x_i^2$ defines the optical angular frequencies,
\begin{equation}
 \omega_x^2=\omega_z^2=\frac{4U_0}{mw_0^2},
 \qquad \omega_y^2=\frac{2U_0}{mz_R^2}.
 \label{eq:trap}
\end{equation}
Combining the optical potential with $m\Phi$ gives the common Newtonian Hamiltonian in the quadratic approximation,
\[
 H_{\mathrm{Newt}}^{\mathrm{trap}}\simeq
 \frac{\bm p^2}{2m}+m\Phi_{\rm ref}-U_0+mgz
 +\frac{m}{2}\sum_i\Omega_i^2 x_i^2,
 \qquad \Omega_i^2=\omega_i^2+Q_i .
\]
Here $\bm p=-\mathrm{i}\hbar\nabla$ is the local momentum operator, and $\Omega_i$ are the angular frequencies including the tidal curvature. For a stable minimum all three $\Omega_i^2$ are positive. The horizontal equilibrium coordinates are zero. The vertical equilibrium follows by completing the square,
\[
 mgz+\frac{m}{2}\Omega_z^2z^2
 =\frac{m}{2}\Omega_z^2\left(z+\frac{g}{\Omega_z^2}\right)^2
 -\frac{mg^2}{2\Omega_z^2}.
\]
The vertical equilibrium coordinate is therefore $x_{0z}=-g/\Omega_z^2$. The low-lying oscillator states centered there approximate the motion localized near the minimum of the smooth trap.

The common optical potential in Eq.~\eqref{eq:trap_coupling} cancels in the Hamiltonian difference and sets the optical curvature entering the oscillator states $\lvert\bm n\rangle$ through Eq.~\eqref{eq:trap}. Using Eq.~\eqref{eq:tidal} with the constant Hessian in Eq.~\eqref{eq:localPhi} gives the retained 2PN difference
\begin{equation}
 \delta H\equiv[H_D^{\mathrm{trap}}-H_{\KG}^{\mathrm{trap}}]_{2\mathrm{PN}}
 =\frac{\eps^2\hbar^2}{4m^3}\sum_i Q_i p_i^2 .
 \label{eq:trap_difference}
\end{equation}
Conversion to laboratory proper time first changes the frequency difference at order $\eps^3$. The two trapped Hamiltonians share the Newtonian and 1PN terms. For an isolated oscillator level, second-order perturbation theory gives the D--KG energy difference as the diagonal matrix element of $\delta H$ in $\lvert\bm n\rangle=\lvert n_x,n_y,n_z\rangle$:
\[
 \delta E_{\bm n}\equiv[E_{D,\bm n}-E_{\KG,\bm n}]_{2\mathrm{PN}}
 =\langle\bm n|\delta H|\bm n\rangle .
\]
For direction $i$, let $a_i$ and $a_i^\dagger$ be the oscillator lowering and raising operators. The momentum is
\[
 p_i=\mathrm{i}\sqrt{\frac{m\hbar\Omega_i}{2}}\,(a_i^\dagger-a_i).
\]
The momentum-squared expectation value in $\lvert\bm n\rangle$ is
\[
 \frac{\langle p_i^2\rangle_{\bm n}}{2m}
 =\frac{\hbar\Omega_i}{2}\left(n_i+\tfrac12\right).
\]
Substitution in the diagonal matrix element gives the energy difference
\begin{equation}
 \delta E_{\bm n}
 =\frac{\eps^2\hbar^3}{4m^2}\sum_i Q_i\Omega_i\left(n_i+\tfrac12\right).
 \label{eq:motional_level_difference}
\end{equation}

For a transition that adds one quantum along direction $i$, let $\bm e_i$ be the unit increment of $n_i$ and keep the other two occupations fixed. The D minus KG difference of the transition frequencies satisfies
\[
 2\pi\hbar\,\delta\nu_i
 =\delta E_{\bm n+\bm e_i}-\delta E_{\bm n}
 =\frac{\eps^2\hbar^3Q_i\Omega_i}{4m^2}.
\]
The common Newtonian transition frequency is $\nu_i=\Omega_i/(2\pi)$, so the gap is $h\nu_i$ with $h=2\pi\hbar$. The fractional difference is
\begin{equation}
 \frac{\delta\nu_i}{\nu_i}=\frac{\eps^2\hbar^2Q_i}{4m^2}.
 \label{eq:motional_gap_difference}
\end{equation}
The vertical entry $Q_z=-2\mu/R_\oplus^3$ is negative, so the D prescription places the vertical line at a lower frequency than the KG prescription for the same trap.

\subsection{Frequency scale and Ramsey readout}
We use the $^{88}\mathrm{Sr}$ optical tweezer of Giardini et al.~\cite{Giardini2026} as a reference, with trapping wavelength $\lambda=813\,\mathrm{nm}$, beam waist $w_0\simeq1.5\,\mu\mathrm m$ and trap depth $U_0/k_B\simeq360\,\mu\mathrm K$. Here $k_B$ is Boltzmann's constant. For the spherical Earth model, we use
$R_\oplus=6378136.6\,\mathrm{m}$ and
$GM_\oplus=3.986004418\times10^{14}\,\mathrm{m^3\,s^{-2}}$~\cite{IERS2010}. The $^{88}\mathrm{Sr}$ isotope mass is
$m=87.9056125\,\mathrm{u}\simeq1.4597\times10^{-25}\,\mathrm{kg}$~\cite{NISTSr,CODATA2022}. These Earth parameters give
$q=GM_\oplus/R_\oplus^3\simeq1.53623\times10^{-6}\,\mathrm{s^{-2}}$
and $Q_z=-2q$. The optical and tidal curvatures in
Eqs.~\eqref{eq:trap} and~\eqref{eq:localPhi} then give the
Newtonian vertical frequency
\[
 \nu_z=\frac{\Omega_z}{2\pi}\simeq39\,\mathrm{kHz},
 \qquad \Omega_z^2=\omega_z^2+Q_z.
\]
Equation~\eqref{eq:motional_gap_difference} then gives the D minus KG frequency difference $\delta\nu_z=\nu_{D,z}-\nu_{\KG,z}$ through 2PN order,
\begin{equation}
 \delta\nu_z=\nu_z\frac{\eps^2\hbar^2Q_z}{4m^2}
 \simeq-1.9\times10^{-54}\,\mathrm{Hz},
 \qquad \frac{\delta\nu_z}{\nu_z}\simeq-5.0\times10^{-59}.
 \label{eq:Srshift}
\end{equation}

In an ideal Ramsey measurement, a first pulse prepares a coherent superposition of two adjacent vertical vibrational states~\cite{Ramsey1950}. The atom evolves in the trap without driving for a time $T$, and a second pulse with a controlled reference phase converts the accumulated relative phase into a population difference. Repeated measurements of the final vibrational occupations then give the Ramsey signal.
During free evolution for laboratory time $T$, each level amplitude acquires a phase factor $e^{-\mathrm i E_{X,k}T/\hbar}$ with $k=n,n+1$. The ratio of the upper and lower phase factors gives the accumulated relative phase
\[
 \phi_X(T)=-\frac{E_{X,n+1}-E_{X,n}}{\hbar}T
          =-2\pi\nu_{X,z}T.
\]
The transverse occupations are held fixed. Subtracting the two prescriptions' phases gives
\[
 |\phi_D-\phi_{\KG}|=2\pi|\delta\nu_z|T.
\]
For an interrogation time equal to the reported vacuum-limited lifetime $T=425\,\mathrm s$~\cite{Giardini2026}, the phase difference reaches $5.2\times10^{-51}\,\mathrm{rad}$ under preserved motional coherence.

\section{Conclusions}\label{sec:discussion}

In this work, we have compared two single-particle quantum prescriptions
in the static Schwarzschild exterior. The D prescription
quantizes the positive classical Hamiltonian after minimal coupling,
whereas the KG construction minimally couples the
covariant wave equation of the flat-spacetime squared-energy operator
before taking the positive spectral root. In the common radial
representation, their squared energies place $RNR$ and $\{R^2,N\}/2$
between the outer factors of $P$, yielding the exact double-commutator
difference in Eq.~\eqref{eq:orders_Cdiff}. The two Hamiltonians agree
through 1PN order and first differ locally at 2PN order through the
tidal momentum coupling in Eq.~\eqref{eq:tidal}. This difference is
purely quantum, vanishes in the flat limit and for a uniform linear
Newtonian potential, and therefore reflects genuine gravitational tidal
gradients rather than uniform-acceleration ordering effects. Within the
symmetric ordering family $T_\alpha$, varying the lapse ordering adds
only a multiplication potential at 2PN order, so no fixed member
reproduces the KG squared-energy operator through this order
(Section~\ref{sec:tidal_difference}).

For the midpoint prescription $T_D$, Eq.~\eqref{eq:gap} yields 2PN
spectral gaps that differ from KG for some pairs of matched bound levels
on the radial half-line in a fixed angular sector. At fixed physical
parameters, time normalization and reflecting Dirichlet boundaries,
these gap shifts rule out unitary equivalence up to a common energy
shift for the retained radial spectral series (Eq.~\eqref{eq:nounitary}).
Unexpanded finite-cavity Galerkin spectra provide an independent
numerical verification of the analytical 2PN coefficients
(Section~\ref{sec:numerical}). More generally, we have shown in
Appendix~\ref{app:pn_orders} that at fixed lapse expansion, the $j$PN
operator differences ($j\geq2$) require kinetic coefficients only
through $K_{j-2}$, and for matched simple levels the $j$PN spectral
differences also require state corrections only through order $j-2$.

As a physical benchmark, adding the same calibrated optical potential to
both generators (Eq.~\eqref{eq:trap_coupling}) produces a vertical
motional frequency difference
$\nu_{D,z}-\nu_{\KG,z}\simeq-1.9\times10^{-54}\,\mathrm{Hz}$ in the
reference $^{88}\mathrm{Sr}$ optical tweezer (Eq.~\eqref{eq:Srshift}),
corresponding to an ideal Ramsey phase difference of
$5.2\times10^{-51}\,\mathrm{rad}$ over $T=425\,\mathrm{s}$. In
terrestrial weak-field atomic platforms, this 2PN ordering discrepancy
lies far below current experimental sensitivity. Consequently, for such
trapped-atom observables, the two prescriptions are indistinguishable at
present laboratory precision, even though they remain spectrally
inequivalent in curved spacetime.

Several extensions follow naturally from the framework established here.
The general order bound and the 3PN formula in
Eq.~\eqref{eq:orders_energy_3pn} allow a direct evaluation of 3PN
spectral differences using only $K_1$ and the common 1PN state
correction. Extending the comparison from static Schwarzschild slices to
stationary spacetimes with nonzero shift, such as the Kerr exterior,
would test how frame dragging interacts with lapse-root ordering. It
would also be worthwhile to examine alternative self-adjoint boundary
conditions, spinning particles and covariant electromagnetic couplings
for the trapping potential beyond the additive model in
Eq.~\eqref{eq:trap_coupling}.

\section*{Acknowledgements}
This work is supported by the National Key Research and Development Program of China (No.~2021YFC2203001) and the National Natural Science Foundation of China (No.~12475049).
\section*{Data availability statement}
All data that support the findings of this study are included within the article, and the verification scripts are available from the corresponding author upon reasonable request.
\appendix
\section{Expanded radial coefficients}\label{app:coeff}

The KG product and its normalization are given in Eqs.~\eqref{eq:kg_sample_product}, \eqref{eq:kg_sample_positive_product} and~\eqref{eq:normKG}.
Expanding $P=f^{1/4}=1+\eps\Phi/2-3\eps^2\Phi^2/8+\cdots$ and $N=1+\eps\Phi-\eps^2\Phi^2/2+\eps^3\Phi^3/2+\cdots$ in $PR_{\mathrm{rad}}P$, together with
\begin{equation}
 R_{\mathrm{rad}}=mc^2+\frac{K_{\mathrm{rad}}}{2m}-\frac{\eps K_{\mathrm{rad}}^2}{8m^3}
                         +\frac{\eps^2 K_{\mathrm{rad}}^3}{16m^5}+\cdots,
\end{equation}
gives Eqs.~\eqref{eq:H1} and~\eqref{eq:H2D}. The rest-energy term requires the $\eps^3$ term of $N$ to obtain the $\eps^2$ rest-subtracted Hamiltonian. Expanding the common root in Eq.~\eqref{eq:Croot} with Eq.~\eqref{eq:Cexact} gives the KG coefficient in Eq.~\eqref{eq:H2KG}.

For direct implementation, write $H_{2,\KG}u=\sum_{j=0}^6b_ju^{(j)}$ and
$H_{2,D}u=\sum_{j=0}^6(b_j+d_j)u^{(j)}$, where $u^{(j)}$ denotes the $j$th derivative of $u$ with respect to $r$. The coefficients are for $H_2$, before scaling by $c^{-4}$:
\begin{align}
 b_6={}&-\frac{\hbar^6}{16m^5},\qquad b_5=0,\nonumber\\
 b_4={}&\frac{3L\hbar^6}{16m^5r^2}+\frac{5\hbar^4\mu}{8m^3r},\nonumber\\
 b_3={}&-\frac{3L\hbar^6}{4m^5r^3}-\frac{5\hbar^4\mu}{4m^3r^2},\nonumber\\
 b_2={}&-\frac{3\hbar^6L(L-14)}{16m^5r^4}
         -\frac{3\hbar^4\mu(L-3)}{4m^3r^3}-\frac{3\hbar^2\mu^2}{4mr^2},\nonumber\\
 b_1={}&\frac{3\hbar^6L(L-8)}{4m^5r^5}
         +\frac{3\hbar^4\mu(3L-4)}{4m^3r^4}+\frac{3\hbar^2\mu^2}{2mr^3},\nonumber\\
 b_0={}&\frac{\hbar^6L(L^2-26L+120)}{16m^5r^6}
         +\frac{\hbar^4\mu(L^2-32L+24)}{8m^3r^5}-\frac{\hbar^2\mu^2(L+6)}{4mr^4}-\frac{m\mu^3}{2r^3},\label{eq:expandedKG}\\
 d_2={}&\frac{\hbar^4\mu}{2m^3r^3},\quad
 d_1=-\frac{3\hbar^4\mu}{2m^3r^4},\quad
 d_0=\frac{\hbar^4\mu(L+6)}{4m^3r^5},\quad d_3=\cdots=d_6=0.
 \label{eq:expandedD}
\end{align}
The local difference in the interior of the radial interval is
\[
 (H_{2,D}-H_{2,\KG})u=d_2(r)u''(r)+d_1(r)u'(r)+d_0(r)u(r).
\]
In the flat limit $\mu=GM=0$, $d_0=d_1=d_2=0$ and the 2PN difference vanishes.
\section{Isotropic coordinates and the 1PN check}\label{app:isotropic}

The radial coordinate $r$ is the areal radius, while an isotropic radius $s$ describes the spatial metric as~\cite{PoissonWill2014}
\begin{equation}
 h_{ij}\dd x^i\dd x^j
 =\left(1+\frac{\mu}{2c^2s}\right)^4
       (\dd s^2+s^2\dd\Omega^2).
 \label{eq:isotropic}
\end{equation}
Matching the angular and radial terms, with $s/r\to1$ at infinity, gives
\begin{equation}
 \frac{\dd r}{r\sqrt f}=\frac{\dd s}{s},\qquad
 r=s\left(1+\frac{\mu}{2c^2s}\right)^2
   =s+\frac{\mu}{c^2}+O(c^{-4}).
 \label{eq:radii}
\end{equation}
On the exterior branch, $s>\mu/(2c^2)$, the lapse follows by substituting the radius relation into $N=\sqrt f$:
\begin{equation}
 N=\sqrt{1-\frac{2\mu}{c^2s(1+\mu/(2c^2s))^2}}
  =\frac{1-\mu/(2c^2s)}{1+\mu/(2c^2s)},
 \label{eq:isotropic_lapse}
\end{equation}
with the same time coordinate and normalization.
For a given geometry, the wall $r=R_0$ is represented in isotropic coordinates by $s=s_0$, where
\begin{equation}
 R_0=s_0\left(1+\frac{\mu}{2c^2s_0}\right)^2,
 \qquad s_0=R_0-\frac{\mu}{c^2}
               -\frac{\mu^2}{4c^4R_0}+O(c^{-6}).
 \label{eq:wall_transport}
\end{equation}
A coordinate change transports the boundary condition together with the wavefunction, measure and operators.

For the vacuum PPN check we use isotropic Cartesian coordinates with radial coordinate $s$ and potential $\Phi_{\rm iso}=-\mu/s$. Differentiating Eq.~\eqref{eq:radii} gives
\begin{equation}
 \frac{\dd r}{\dd s}=1+O(\eps^2).
 \label{eq:agreement_sample_coordinates}
\end{equation}
Preserving the radial scalar product gives
\begin{equation}
 \begin{aligned}
 u_s(s)&=\sqrt{\frac{\dd r}{\dd s}}\,u_r(r(s))=u_r(s)+\mu\eps\,\partial_su_r(s)+O(\eps^2).
 \end{aligned}
 \label{eq:radial_coordinate_amplitude}
\end{equation}
Thus the map between the corresponding radial spaces expands as
\begin{equation}
 U(\eps)=1+\mu\eps\partial_s+O(\eps^2),\qquad
 U(\eps)^{-1}=1-\mu\eps\partial_s+O(\eps^2).
 \label{eq:radial_coordinate_map}
\end{equation}
For $H_{\mathrm{rad}}^r(\eps)=H_0+\eps H_1^r+\cdots$, the transformed Hamiltonian satisfies
\begin{align}
 H_{\mathrm{rad}}^s(\eps)&=U(\eps) H_{\mathrm{rad}}^r(\eps)U(\eps)^{-1}\nonumber\\
 &=(1+\mu\eps\partial_s)(H_0+\eps H_1^r)
      (1-\mu\eps\partial_s)+O(\eps^2)\nonumber\\
 &=H_0(s)+\eps\left[
     H_1^r\big|_{r=s}+\mu[\partial_s,H_0(s)]
   \right]+O(\eps^2),
 \label{eq:radial_coordinate_H}
\end{align}
where
\begin{equation}
 H_0(s)=-\frac{\hbar^2}{2m}\partial_s^2
       +\frac{\hbar^2L}{2ms^2}-\frac{m\mu}{s}.
 \label{eq:isotropic_H0}
\end{equation}
The 1PN coefficient is therefore
\begin{align}
 H_1^s
 &=H_1^r\big|_{r=s}+\mu[\partial_s,H_0(s)]\nonumber\\
 &=H_1^r\big|_{r=s}
       -\frac{\hbar^2\mu L}{ms^3}+\frac{m\mu^2}{s^2}\nonumber\\
 &=-\frac{K_0(s)^2}{8m^3}
    +\frac{3\hbar^2\mu}{2m}\partial_s(s^{-1}\partial_s)
    +\frac{3\hbar^2\mu(1-L)}{2ms^3}
    +\frac{m\mu^2}{2s^2},
 \label{eq:agreement_sample_isoradial}
\end{align}
where $K_0(s)=-\hbar^2\partial_s^2+\hbar^2L/s^2$.

In the flat Cartesian measure, Weyl ordering gives~\cite{Zworski2012}
\begin{equation}
 \Op(\Phi_{\rm iso}p^2)
 =-\hbar^2\left(\Phi_{\rm iso}\Delta
       +\nabla\Phi_{\rm iso}\cdot\nabla
       +\frac14\Delta\Phi_{\rm iso}\right).
 \label{eq:agreement_sample_Weyl}
\end{equation}
Outside the source $\Delta\Phi_{\rm iso}=0$. Acting on $u_s(s)Y_{\ell q}/s$ and extracting its radial amplitude reduces this expression to $\hbar^2\mu[\partial_s(s^{-1}\partial_s)+(1-L)/s^3]$. Hence the radial coefficient~\eqref{eq:agreement_sample_isoradial} is precisely the fixed-angular-momentum representation of
\begin{equation}
 H_1^{\rm iso}
 =\Op\left(-\frac{p^4}{8m^3}
       +\frac{3\Phi_{\rm iso}p^2}{2m}
       +\frac{m\Phi_{\rm iso}^2}{2}\right),
 \label{eq:SG1PN}
\end{equation}
which agrees with the static, general-relativistic specialization of the normalized PPN Hamiltonian in Ref.~\cite{SchwartzGiulini2019}, Sec.~3.4.
The coordinate change of Eq.~\eqref{eq:radii} also sends each fixed areal wall $r=R_i$ to $s_i=R_i-\mu\eps+O(\eps^2)$, with $u_s(s_i)=0$. Together with the transformation~\eqref{eq:radial_coordinate_amplitude}, this preserves the original spectral problem.

\section{Kinetic and state coefficients at higher PN orders}\label{app:pn_orders}

We expand the squared-energy relations of Section~\ref{sec:squared_energy}
to arbitrary PN order at fixed areal radii~\cite{SchwartzThesis2020,GiuliniGrossardtSchwartz2023}.
We suppress the radial subscript and write $[\eps^j]$ for the coefficient of $\eps^j$.

\subsection{Kinetic coefficients in the Hamiltonians and their difference}\label{app:pn_kinetic}
We evaluate the local differential-operator coefficients on the same test space as in Section~\ref{sec:squared_energy}.
Equation~\eqref{eq:orders_CKG} gives the dependence of $C_\KG$ on $K$. In the fixed areal-radius expansion, $N^2=1+2\eps\Phi$ holds exactly.
If $N=\sum_{a\geq0}\eps^aN_a$ and $P=\sum_{a\geq0}\eps^aP_a$,
with $N_0=P_0=1$, the coefficient of order $j\geq0$ is
\begin{equation}
 C_{\KG,j}=m\Phi\,\delta_{j0}
  +\frac{1}{4m}\sum_{a+b+d+k=j}P_a\{K_k,N_b\}P_d ,
 \qquad a,b,d,k\geq0 .
 \label{eq:orders_CKG_coeff}
\end{equation}
The term with $k=j$ is $K_j/(2m)$. In this representation the
exact differential expression in Eq.~\eqref{eq:Cexact} also gives
the finite polynomial $C_\KG=H_0+\eps C_{\KG,1}+\eps^2C_{\KG,2}$ of
Eqs.~\eqref{eq:kg_sample_Bexpansion} and~\eqref{eq:Ccoeff}.
Thus the lapse and kinetic contributions in
Eq.~\eqref{eq:orders_CKG_coeff} cancel identically for $j\geq3$.

We write $H_X=\sum_{j\geq0}\eps^j H_{X,j}$ and set
$\delta H_j=H_{D,j}-H_{\KG,j}$ and
$\delta C_j=[\eps^j](C_D-C_\KG)$. Products in the coefficient
formulas of this subsection refer to the local expressions.
The binomial expansion of $S$ is
\begin{equation}
 S=\sum_{q\geq1}\binom{1/2}{q}
                 \frac{\eps^{q-1}}{m^{2q-1}}K^q .
 \label{eq:orders_S}
\end{equation}
For $S_i=[\eps^i]S$, the $q=1$ term gives $K_i/(2m)$.
For $q\geq2$ the coefficient uses only $K_0,\ldots,K_{i-1}$.
In particular $S_0=K_0/(2m)$ and
$S_1=K_1/(2m)-K_0^2/(8m^3)$.
The identity $H_D=m(N-1)/\eps+PSP$ then gives
\begin{equation}
 H_{D,j}=mN_{j+1}+\sum_{a+i+b=j}P_aS_iP_b ,
 \qquad a,i,b\geq0 .
 \label{eq:orders_HD_coeff}
\end{equation}

The relation between $C_X$ and $H_X$ in Section~\ref{sec:squared_energy} gives, for $j\geq1$,
\begin{equation}
 H_{X,j}=C_{X,j}
         -\frac{1}{2m}\sum_{a+b=j-1}H_{X,a}H_{X,b},
 \qquad a,b\geq0 ,
 \label{eq:orders_H_coeff}
\end{equation}
with $H_{X,0}=C_{X,0}=H_0$.
For D, the term with $a=b=0$ and $i=j$ in Eq.~\eqref{eq:orders_HD_coeff} gives $S_j$, which contains $K_j/(2m)$ by Eq.~\eqref{eq:orders_S}. All other terms in the sum use lower kinetic coefficients. For KG, the term with $a=b=d=0$ and $k=j$ in Eq.~\eqref{eq:orders_CKG_coeff} gives $K_j/(2m)$ in $C_{\KG,j}$. The products in Eq.~\eqref{eq:orders_H_coeff} involve only lower orders of $H_\KG$ and thus use kinetic coefficients through at most $K_{j-1}$.
In the general expansion, $K_j$ is the highest kinetic coefficient that enters $H_{X,j}$ in either prescription. In this Schwarzschild representation, $C_{\KG,j}=0$ for $j\geq3$. The recursion in Eq.~\eqref{eq:orders_H_coeff} then generates all $H_{\KG,j}$ from $H_0$, $C_{\KG,1}$ and $C_{\KG,2}$.

We note that, at fixed lapse expansion, the $j$PN differences
$\delta C_j$ and $\delta H_j$ for $j\geq2$ require kinetic
coefficients only through $K_{j-2}$. Expanding
Eq.~\eqref{eq:orders_Cdiff} and using $[S,N_0]=0$ gives
\begin{equation}
 \delta C_j=-\frac{1}{4m}
 \sum_{\substack{a+b+d+i+k=j-1\\ b\geq1,\ a,d,i,k\geq0}}
      P_a[S_i,[S_k,N_b]]P_d .
 \label{eq:orders_Cdiff_coeff}
\end{equation}
Hence $i,k\leq j-2$. The same order count holds for $c_D-c_\KG$ by
Eq.~\eqref{eq:squared_energy_norm_difference}, since its commutator also starts at order $\eps$. At $j=2$ the only term
has $i=k=a=d=0$ and $b=1$, with $N_1=\Phi$, reproducing the
local difference in Eq.~\eqref{eq:delta}. Subtracting
Eq.~\eqref{eq:orders_H_coeff} for the two prescriptions gives
\begin{equation}
 \delta H_j=\delta C_j-\frac{1}{2m}\sum_{a+b=j-1}
       \bigl(\delta H_a H_{D,b}+H_{\KG,a}\delta H_b\bigr).
 \label{eq:orders_Hdiff_coeff}
\end{equation}
Since $\delta H_0=\delta H_1=0$, each nonzero product has a
difference coefficient of index at least two and a single-Hamiltonian
coefficient of index at most $j-3$. The difference coefficient in
each product also has index less than $j$. This proves the bound
by induction. Thus $K_0$, $K_1$ and $K_2$ are the highest kinetic coefficients that can enter the difference at 2PN, 3PN and 4PN, respectively. These indices give upper bounds on the required kinetic input, while individual coefficients may simplify further.

We now expand the symmetric ordering family of
Section~\ref{sec:positive} through 2PN. For fixed $\alpha$, set
$b=\alpha-1/2$ and $L_N=\log N$. Since $N^s=\exp(sL_N)$, the
conjugated operator $F(s)=N^sRN^{-s}$ satisfies
$F'(s)=[L_N,F(s)]$ and $F''(0)=[L_N,[L_N,R]]$.
Removing the two outer factors of $P$ gives
\[
\begin{aligned}
    P^{-1}(T_\alpha-T_D)P^{-1}
    &=\frac{F(b)+F(-b)}{2}-R=\frac{b^2}{2}[L_N,[L_N,R]]+O(c^{-8}).
\end{aligned}
\]
The average cancels the odd Taylor terms. The next even term
contains four factors of $L_N=O(c^{-2})$, and the rest energy
$mc^2$ commutes with $L_N$. Using $L_N=\Phi/c^2+O(c^{-4})$ and
$R=mc^2+K_0/(2m)+O(c^{-2})$, we obtain
\[
    T_\alpha-T_D
    =\frac{b^2}{4mc^4}[\Phi,[\Phi,K_0]]+O(c^{-6}).
\]
Here the outer factors $P=1+O(c^{-2})$ affect the difference
only from order $c^{-6}$ onward. In Newtonian Cartesian
coordinates, $K_0=-\hbar^2\Delta$. The product rule gives, on
interior test functions $u$,
\[
\begin{aligned}[c]
    [\Phi,K_0]u
    &=\hbar^2\bigl((\Delta\Phi)u+2\nabla\Phi\cdot\nabla u\bigr),\\
    [\Phi,[\Phi,K_0]]u
    &=2\hbar^2\bigl(\Phi\nabla\Phi\cdot\nabla u
      -\nabla\Phi\cdot\nabla(\Phi u)\bigr)=-2\hbar^2|\nabla\Phi|^2u.
\end{aligned}
\]
This gives the multiplication potential quoted in
Section~\ref{sec:tidal_difference}. The nonzero second-order
differential term in Eq.~\eqref{eq:tidal} therefore survives in
$T_\alpha-T_{\KG}$ for every fixed $\alpha$. By
Eq.~\eqref{eq:agreement_sample_Cdefinition}, the same 2PN
difference occurs in the normalized squared energies.

\subsection{Energy differences on matched spectral branches}\label{app:pn_spectral}
Let $n$ label a simple isolated level of $H_0$ in a fixed radial sector, as in Section~\ref{sec:spectrum}. We work on matched Dirichlet branches whose eigenvalues and eigenstates admit PN expansions through the order considered, with well-defined coefficient pairings.
The spectral calculation uses $c_X$ from Section~\ref{sec:spectrum}.
Write the expansions as
\begin{equation}
 u_{X,n}=\sum_{a\geq0}\eps^a u_{X,n}^{(a)},\qquad
 u_{D,n}^{(0)}=u_{\KG,n}^{(0)}=u_n,
 \label{eq:orders_state}
\end{equation}
where $\langle u_n,u_n\rangle=1$ and $\langle u_n,u_{X,n}^{(a)}\rangle=0$
for $a\geq1$. This intermediate normalization fixes the component
along $u_n$ at each order~\cite{Kato1995,ReedSimon1978}.

Testing each squared-energy eigenvalue equation with the state
from the other branch gives, with the scalar product linear in
its second argument,
\begin{equation}
 (\lambda_{D,n}-\lambda_{\KG,n})\langle u_{\KG,n},u_{D,n}\rangle
   =(c_D-c_\KG)[u_{\KG,n},u_{D,n}] .
 \label{eq:orders_cross_form}
\end{equation}
The scalar relation~\eqref{eq:squared_energy_eigenvalue} then gives the general formula
\begin{equation}
 E_{D,n}-E_{\KG,n}=
 \frac{(c_D-c_\KG)[u_{\KG,n},u_{D,n}]}
      {\langle u_{\KG,n},u_{D,n}\rangle
       \bigl[1+\eps(E_{D,n}+E_{\KG,n})/(2m)\bigr]} .
 \label{eq:orders_energy_diff}
\end{equation}
Both factors in the denominator have constant term one.
Normalizing and polarizing the exact pairing in Eq.~\eqref{eq:squared_energy_norm_difference},
with $R=S+mc^2$, gives the coefficient $\delta c_r=[\eps^r](c_D-c_\KG)$ for $r\geq2$,
\begin{equation}
 \delta c_r[u,v]=-\frac{1}{4m}
 \sum_{\substack{a+b+d+i+k=r-1\\b\geq1,\ a,d,i,k\geq0}}
 \bigl(\langle S_iP_a u,[S_k,N_b]P_d v\rangle
       +\langle [S_k,N_b]P_a u,S_iP_d v\rangle\bigr).
 \label{eq:orders_weak_coeff}
\end{equation}
The numerator coefficient at order $j\geq2$ is therefore
\begin{equation}
 \sum_{\substack{r+a+b=j\\r\geq2,\ a,b\geq0}}
 \delta c_r[u_{\KG,n}^{(a)},u_{D,n}^{(b)}] .
 \label{eq:orders_spectral_numerator}
\end{equation}
Taking $[\eps^j]$ in Eq.~\eqref{eq:orders_energy_diff} therefore
requires state corrections at most through order $j-2$. Through
that order the overlap is determined by the same state coefficients,
and the energy factor uses $E_{X,n}$ only through order $j-3$
when $j\geq3$.

Insert $c_X=\sum_{b\geq0}\eps^b c_{X,b}$, the state expansion in
Eq.~\eqref{eq:orders_state}, and $\lambda_{X,n}=\sum_{b\geq0}\eps^b\lambda_{X,n}^{(b)}$
into $c_X[v,u_{X,n}]=\lambda_{X,n}\langle v,u_{X,n}\rangle$. The coefficient of
$\eps^a$ for $a\geq1$ obeys
\[
 \sum_{b=0}^a c_{X,b}[v,u_{X,n}^{(a-b)}]
 =\sum_{b=0}^a\lambda_{X,n}^{(b)}\langle v,u_{X,n}^{(a-b)}\rangle .
\]
Taking $v=u_n$ and using $\langle u_n,u_{X,n}^{(k)}\rangle=0$ for $k\geq1$, we obtain
\begin{equation}
 \lambda_{X,n}^{(a)}
 =\sum_{b=1}^{a}c_{X,b}[u_n,u_{X,n}^{(a-b)}] .
 \label{eq:orders_lambda_recursion}
\end{equation}
Using $c_{X,0}[v,w]=\langle H_0v,w\rangle$ and
$\lambda_{X,n}^{(0)}=e_n^{(0)}$, the $b=0$ terms give
\[
 \bigl\langle(H_0-e_n^{(0)})v,u_{X,n}^{(a)}\bigr\rangle
 =-\sum_{b=1}^a\left(
 c_{X,b}[v,u_{X,n}^{(a-b)}]
 -\lambda_{X,n}^{(b)}\langle v,u_{X,n}^{(a-b)}\rangle
 \right).
\]
For $v\perp u_n$, the term $\lambda_{X,n}^{(a)}\langle v,u_n\rangle$ vanishes.
The inverse of $H_0-e_n^{(0)}$ on $u_n^\perp$ then determines $u_{X,n}^{(a)}$
from the lower-order state and eigenvalue coefficients and
$c_{X,1},\ldots,c_{X,a}$~\cite{Kato1995,ReedSimon1978}.
Equations~\eqref{eq:orders_CKG_coeff}
and~\eqref{eq:orders_Cdiff_coeff} therefore show that
the $j$PN energy difference requires kinetic
coefficients through at most $K_{j-2}$, at fixed lapse expansion.
From 4PN onward the state corrections entering the formula can
come from different branches, since their coefficients can first
differ at 2PN.

At 2PN, Eq.~\eqref{eq:orders_energy_diff} reduces to
$\delta c_2[u_n,u_n]$, giving the 2PN coefficient in
Eq.~\eqref{eq:wall_alt_coeff}.
At 3PN, write $v_n=u_{D,n}^{(1)}=u_{\KG,n}^{(1)}$ and
$\delta e_n^{(j)}=[\eps^j](E_{D,n}-E_{\KG,n})$. The formula gives
\begin{equation}
 \delta e_n^{(3)}
 =\delta c_3[u_n,u_n]
   +2\operatorname{Re}\delta c_2[u_n,v_n]
   -\frac{e_n^{(0)}}{m}\,\delta c_2[u_n,u_n] .
 \label{eq:orders_energy_3pn}
\end{equation}
The last term comes from converting the squared-energy eigenvalues
to the rest-subtracted energies. The 3PN expression needs $K_1$
through $\delta c_3$ and through the common first-order state
correction.

\bibliographystyle{unsrt}
\bibliography{references}
\end{document}